\documentclass[conference]{IEEEtran}
\IEEEoverridecommandlockouts

\mathchardef\mhyphen="2D

\newtheorem{definition}{\textbf{Definition}}

\newcommand{\adv}{\ensuremath{\mathcal{A}}}
\newcommand{\bdv}{\ensuremath{\mathcal{B}}}
\newcommand{\linkgame}[2]{\hyperref[#1]{G#2}}

\newcounter{Bdversary}

\newcommand\orcidicon[1]{\href{https://orcid.org/#1}{\mbox{\scalerel*{
\begin{tikzpicture}[yscale=-1,transform shape]
\pic{orcidlogo};
\end{tikzpicture}
}{|}}}}

\newcommand*{\addFileDependency}[1]{% argument=file name and extension
  \typeout{(#1)}% latexmk will find this if $recorder=0 (however, in that case, it will ignore #1 if it is a .aux or .pdf file etc and it exists! if it doesn't exist, it will appear in the list of dependents regardless)

  \IfFileExists{#1}{}{\typeout{No file #1.}}% latexmk will find this message if #1 doesn't exist (yet)
}

\usepackage{cite}
\usepackage{amsmath,amssymb,amsfonts}
\usepackage{algorithmicx}
\usepackage{booktabs}
\usepackage{graphicx}
\usepackage{textcomp}
\usepackage{xcolor}
\usepackage[noend]{algpseudocode}
\usepackage{dashbox}
\usepackage{adjustbox}
\usepackage{mdframed}
\usepackage{todonotes}
\usepackage{tikz}
\usepackage{hyperref}
\usepackage{subcaption}
\usepackage{algorithm}
\usepackage{algpseudocode}
\usepackage{mathtools}
\usepackage{enumitem}
\usetikzlibrary{positioning, shapes.geometric, arrows.meta, fit, calc}
\usepackage{fancyhdr}
\usepackage [
    n, % or lambda 
    advantage,
    operators,
    sets,
    adversary,
    landau,
    probability,
    notions,
    logic,
    ff, 
    mm,
    primitives,
    events,
    complexity,
    oracles,
    asymptotics,
    keys
]{cryptocode}
\usepackage{threeparttable}
\usepackage[symbol]{footmisc}
\def\BibTeX{{\rm B\kern-.05em{\sc i\kern-.025em b}\kern-.08em
    T\kern-.1667em\lower.7ex\hbox{E}\kern-.125emX}}
\begin{document}
\renewcommand{\headrulewidth}{0pt}
\fancypagestyle{title}{%
  \fancyhf{} % 清除默认页眉页脚
  \fancyhead[C]{This paper has been accepted by IEEE International Symposium on Hardware Oriented Security and Trust (HOST) 2024} % 在'title'页的左侧页眉处添加文本
  \renewcommand{\headrulewidth}{0pt} % 隐藏页眉下方的横线
  % \rfoot{Page \thepage} % 在'title'页的右侧页脚处添加页码
}

\title{PhenoAuth: A Novel PUF-Phenotype-based Authentication Protocol for IoT Devices\\
}

% PUF-Phenotype-based Authentication Protocol of IoT devices.
% \and
% \and
% \and
% \and
% \and

\author{Fei Hongming\textsuperscript{1}, Owen Millwood\textsuperscript{2}, Prosanta Gope\textsuperscript{2}, Jack Miskelly\textsuperscript{3},  Biplab Sikdar\textsuperscript{1}\\
\textsuperscript{1}Department of Electrical and Computer Engineering, The National University of Singapore, Singapore\\
fei.hongming@u.nus.edu, bsikdar@nus.edu.sg\\
\textsuperscript{2}Department of Computer Science, The University of Sheffield, Sheffield, UK\\
\{ojwmillwood1, p.gope\}@sheffield.ac.uk\\
\textsuperscript{3}Centre for Secure Information Technologies, Queen's University Belfast, United Kingdom\\
j.miskelly@qub.ac.uk\\
}

\maketitle
\thispagestyle{title}

\begin{abstract}
Physical Unclonable Functions (PUFs) have been shown to be a highly promising solution for enabling high security systems tailored for low-power devices. Commonly, PUFs are utilised to generate cryptographic keys on-the-fly, replacing the need to store keys in vulnerable, non-volatile memories. Due to the physical nature of PUFs, environmental variations cause noise, manifesting themselves as errors which are apparent in the initial PUF measurements. This necessitates expensive active error correction techniques which can run counter to the goal of lightweight security. ML-based techniques for authenticating noisy PUF measurements were explored as an alternative to error correction techniques, bringing about the concept of a \textit{PUF Phenotype}, where PUF identity is considered as a structure agnostic representation of the PUF, with relevant noise encoding. This work proposes a full noise-tolerant authentication protocol based on the PUF Phenotype concept and methodology for an Internet-of-Things (IoT) network, demonstrating mutual authentication and forward secrecy in a setting suitable for device-to-device communication. Upon conducting security and performance analyses, it is evident that our proposed scheme demonstrates resilience against various attacks compared to the currently existing PUF protocols.

%\textcolor{red}{Would be good to add a sentence or two to talk about the results and how they represent an advantage over existing techniques.}
\end{abstract}

\section{Introduction}
% Background -- IoT, Authentications, PUFs, 
The Internet of Things (IoT) describes a network where heterogeneous physical devices interconnect with each other so that all devices in it can communicate and exchange data smartly, quickly and safely.
%IoT is defined as such a network and all the devices inside. 
In recent years, IoT applications and devices have developed rapidly, e.g., smart homes\cite{poh2019privhome}, Industry 4.0 \cite{lasi2014industry}, smart grids, and healthcare. It brings significant changes to industrial and societal production and management that allow the monitoring and control of multiple sensors in detail and then precisely manage the operations ranging from data collection and processing to analysis and interventions. However, IoT devices manage, process or store confidential and sensitive data, e.g., users' private data and passwords, and in most cases, are lightweight and have limited computing resources, which raises security concerns.  While numerous authentication protocols have been proposed \cite{Ren2021Group, Gope2018RFID,AysuEnd2End,Zheng2022P2P,yu2016lockdown}, several practical challenges remain inadequately addressed. These include the reliance on a trustworthy third-party verifier or server, the lack of secure and reliable non-volatile memory (NVM), and the resource constraints inherent in group authentication scenarios.

Physical Unclonable Function (PUF) is a secure primitive that provides a streamlined solution for lightweight device authentication. It is rooted in the physical characteristics of devices that are produced during the manufacturing process, which are considered to be ``uncontrolled" even by the manufacturers. PUFs are widely used in lightweight scenarios since they can provide unclonable and ``invasive-resistant" entropy for security applications with a low cost of resources. In general, PUFs use challenges as input and provide responses as the output. The challenges can be viewed as a different stimulus to the hardware, which may control how the device is activated or initialized. The readout of responses takes several clock cycles but do not consume too many resources. Although PUFs can provide lightweight, secure primitives for authentication, it has two significant drawbacks. The first is that PUFs are found to be vulnerable to Machine-Learning Modelling-Attacks (MLMAs); if one adversary with machine-learning capability can collect enough challenge-response pairs (CRPs), then he/she can model the PUF with high accuracy. Thus, the CRPs used for authentication should be hidden from adversaries. Secondly, PUFs' responses contain noise when they are evaluated, which results in unreliability in practical applications. One countermeasure is using error correction algorithms to correct errors between evaluations. However, such methods either need pre-shared or real-time generated helper data, which undergrade the security of PUFs and, thus, the authentication protocols. For example, \cite{delvaux2014attacking} has shown that the adversary can get a non-negligible advantage from learning and manipulating the helper data. Secure and efficient authentication protocols for IoT devices are still under research.

\subsection{Related Work}

In response to such challenges with protocol-level integration of PUFs, various research has investigated extraneous techniques for measuring PUF response authenticity without explicit de-noising steps. Intuitively, machine learning (ML) techniques appeared as a clear candidate for solving the problem. Karimian et al. introduced the first ML-based authentication of PUFs \cite{dram-net}. Their work exploited a deep Convolutional Neural Network (CNN) to classify the origin of noisy DRAM-PUF-derived responses formed as image data. Najafi et al. performed a similar study utilising CNNs to authenticate Dynamic Random Access Memory (DRAM) PUF responses by characterising responses which represent the physical sub-structure of the DRAM cells used to derive the response, enabling this information to be encoded into the model \cite{deeppuf}. Suragani et al. investigated using CNNs to classify noisy Static Random Access Memory (SRAM) PUF responses, even when corrupted with large artefacts in the images \cite{reshmi_sram_cnn}. Millwood et al. introduced the concept of a \textit{PUF Phenotype} as a means to define PUF data as a structure-agnostic representation of the PUF, including all noise sources, as opposed to the previous methods by which structural PUF data is encoded in response to be authenticated \cite{millwood2023puf}. In this work, the authors also utilise a modified CNN to perform classification and authentication across multiple devices.
% \textcolor{red}{We need a sentence here to say why this work is still needed and different since CNNs and PUF Phenotypes have already been proposed/used. Even though this is being discussed in the next subesction, it is good to add here for the sake of completeness}

\subsection{Motivation and Contributions}
% \todo[inline]{note: some things stated here will probably be repetitions, depending on what we say in the related work section. We should do an edit once that section has been written.}

Recently published works have demonstrated that noisy hardware identifying information can be used directly to distinguish between nominally identical devices by treating recognition of the noisy IDs as a computer vision problem. The dominant approach to these identities has been to require a de-noising (i.e., error correction) step, with many protocols requiring bit-perfect correction. In some cases, helper data is also required to filter or how the identifying information is used (e.g., filtering high noise sections). Poorly designed helper data can itself be a source of vulnerability. 

In contrast, the computer vision approach (utilising CNNs) requires no error correction and no helper data because the recognition models are inherently noise-tolerant. In fact, they can use the noise distribution as a recognition feature, considering that only a minority subset of the noise produced by a physical system is truly random in position, magnitude, and distribution. That means the resource cost of authentication is shifted almost entirely to the verifying device. Additionally, while the initial methods for classifying noisy PUF responses with CNNs replaced the helper data problem, they did not explicitly tackle the resource overhead issue, with fully connected deep CNNs also requiring a large amount of device storage to operate. Millwood et al. demonstrated that CNN-based PUF authentication need not require the resource-heavy fully connected layer, significantly reducing the required storage to achieve highly accurate authentication.
However, this approach is not an unqualified improvement in all aspects. So far, it has only been tested on fairly large IDs, larger than typical in comparable systems. More importantly, it has sufficiently different properties from the dominant on-device error correction method such that it cannot be substituted into existing protocols. That, in turn, makes it difficult to compare the practicality of deploying it in a real system or to perform a proper vulnerability analysis. Thus, we lack a set of full protocols for this type of authentication, and as a result, we cannot accurately evaluate the system-level pros and cons or determine which scenarios it may be suitable or unsuitable for. Finally, existing machine learning-based PUF authentication systems \cite{yu2016lockdown} exclusively account for a solitary device within a trained model. Consequently, when contemplating a set of $n$ provers, a designated verifier must retain the $n - 1$  model to manage authentication requests from every other member of the group. Given the resource constraints of an IoT device, this is an unrealistic requirement for on-device authentication.

This paper addresses all the aforementioned issues by proposing PhenoAuth, a complete PUF-Phenotype-based authentication protocol for IoT devices. As it is common in IoT networks to have both device-to-server and device-to-device communication, this protocol allows for both modes. This also allows for comparison with as wide a range of existing protocols as possible. We provide algorithms for ID generation, model training, enrollment and deployment, and device authentication. Other than the additional model training step, these are directly comparable to the stages of a conventional PUF protocol. 

We use the proposed protocol to provide a system-level vulnerability analysis and a comparison with previously proposed protocols. It should be noted this is not intended to be a definitive protocol, rather initial steps aimed at providing a solution for a reasonably generic authentication scenario. The hope is this can be used as a starting point for further analysis, refinement, and application to more specific scenarios where this approach can prove advantageous. 

% \todo[inline]{Fill in contributions - OWEN}
Overall, this work provides the following key contributions:
\begin{enumerate}
    \item A novel PUF-based authentication scheme using the concept of \emph{PUF-Phenotype}, where a group of IoT devices (that may act as both prover and verifier) can authenticate each other. Unlike the existing group-based authentication scheme, devices in our proposed scheme are not required to store any \emph{group key} and can detect any alterations or erasures on the stored data, which means that they achieve the same important property of PUFs, tamper resistance. 
    \item  A single ML-based PUF authentication model that is appropriate for multiple group devices that can run on lightweight IoT devices without a third-party trusted verifier.
    \item A concrete analysis and evaluation are performed on the proposed protocol. Formal analysis based on the sequence of games shows that the protocol can guarantee mutual authentication, privacy, and backward and forward security against multiple attacks based on the Dolev-Yao (DY) adversary model.
\end{enumerate}

%We propose PhenoAuth, a PUF-Phenotype-based authentication protocol for IoT devices, where no error correction methods or any insecure helper data is needed.

%\todo[inline]{Application Scenarios -- IoT}
%In smart home, multiple IoT devices communicate to the gateway or each other. As shown in Figure \ref{fig:smart_home}.

% \begin{figure}
%     \centering
%     \includegraphics[width=\linewidth]{figures/IoT.drawio.png}
%     \caption{IoT Communication Scenario}
%     \label{fig:smart_home}
% \end{figure}

\subsection{Paper organisation}
The rest of the paper is organized as follows. In Section II, we first provide a brief introduction to DRAM PUF and PUF-Phenotype. This section also presents a description of the reliability analysis of DRAM memory PUFs. Section III presents our proposed privacy-preserving PUF-Phenotype-based authentication protocol for IoT devices. A comprehensive formal security proof of the proposed scheme is given in Section IV.  Performance analysis of the proposed protocol is then provided in Section V. Finally, we conclude our paper with concluding remarks in Section VI. 
% The symbols and cryptographic functions used in the proposed scheme are defined in Table I.
\section{Preliminaries}

\subsection{DRAM PUF}
DRAM PUFs (DPUFs) are a type of PUF that utilize the inherent physical imperfections in DRAMs for security purposes, which was first proposed by \cite{tehranipoor2016dram}. Almost all computers, ranging from high-performance ones to low-resource constraint IoT devices, are equipped with DRAM, which shows the portability and wide adaptability of DRAM PUFs.
The memory-PUF referenced in this work is the DRAM PUF used in \cite{millwood2023puf}, which used data collected from DDR3 memory on a commodity system. As noted in that paper, this is only exemplary of memory PUFs in general, and the same techniques should transfer to other PUFs based on memory structures (in particular SRAM PUFs), which have been demonstrated to possess the same properties as regards PUF identity and noise characteristics. However, it should be noted that this has not yet been proven experimentally on PUFs other than the DRAM PUF.  

The proposed protocol uses a fairly large block of noisy PUF data as the primary ID. However, it also requires a smaller de-noised (bit-perfect) key to be used in the transmission of this data. Previous works on memory PUFs have shown that within the cells of memory being used as a PUF, there is a reliability distribution, with a minority of cells exhibiting much higher reliability (i.e., less noise). For example, a large-scale test of DRAM PUFs on chips from multiple manufacturers found that at least 2.67\% of cells had reliability greater than 99\%  before error correction. This behaviour is ideal for this protocol, as these low-noise cells can be targeted for key generation with minimal error correction. The remaining noisy cells can be used without error correction as the main ID. This, is in contrast to how this PUF would be used in a conventional protocol, where either the system is limited to using only this subset of low-noise cells or a significant degree of error correction scaled to the noise level in the worst-case cells is needed.

\subsection{PUF-Phenotype}

As previously mentioned, the concept of PUF-Phenotype was first proposed by Millwood et al. \cite{millwood2023puf}, which describes a detectable expression of PUF measurements (experimentally, a DRAM-PUF) including noise from all sources and their relevant encoding, without explicit features describing the underlying PUF structure. This method was proposed to be analogous to biometric authentication methods such as facial recognition, whereby authentication is achieved through classification of the image of a face, including noisy features such as angle, light, facial defects etc. This method provides some key benefits, as PUF measurements can be taken with minimal post-processing (only conversion to an image format is required) and no error correction is required. It was determined in \cite{millwood2023puf} that PUF Phenotypes of the DRAM-PUF could be authenticated very effectively.  
Generating a PUF Phenotype image consists of the process described in Algorithm \ref{alg:PUF-Phenotype}, whereby raw PUF measurements are taken across varying operating conditions to capture the noise profile of the particular PUF. The measurements are then converted into a two-dimensional matrix, where each individual matrix cell contains an integer in the range [0-255], the output being a grayscale image with each pixel denoting pixel intensity. This image forms the input to train DPAN (DRAM-PUF Authentication Network), the classification CNN based on VGG-16 as demonstrated in \cite{millwood2023puf}. The training process of DPAN is described in Algorithm \ref{alg:DPAN}.

% \todo[inline]{How PUF-Phenotype is produced}

\begin{algorithm}
\caption{PUF-Phenotype Generation Algorithm}
\label{alg:PUF-Phenotype}
\begin{algorithmic}[1]
\State \textbf{Dataset Generation}
\State \textbf{Input:} $C \in \{c_0, c_1, \ldots, c_n\}$: Challenge Data
\State \textbf{Output:} $X$: Phenotype dataset
\State \textbf{Data:} $P$: Environmental parameters (Temp., Voltage)
\State $m$: Number of devices
\State $n$: Number of challenges
\For{$i \gets 0 \text{ to } P$}
    \For{$j \gets 0 \text{ to } m$}
        \For{$k \gets 0 \text{ to } n$}
            \State Write challenge pattern from the start location
            \State Set $t_{RCD}$ to $0$
            \State $R_{ijk} \gets$ Perform read operation on DRAM block
            \State $x \gets \text{IMGEN}(R_{ijk})$
            \State \textit{/* $x$: PUF Phenotype */}
            \State Assign device label to $T$
            \State Store $x$ in $X$
        \EndFor
    \EndFor
\EndFor
\State \textbf{return} $X$
\end{algorithmic}
\end{algorithm}

\begin{algorithm}
\caption{Model Training}
\label{alg:DPAN}
\begin{algorithmic}[1]
\State \textbf{Input:} $X$: PUF Phenotype dataset
\State \textbf{Output:} $DPAN$: Trained model; 
        \State \quad\quad\quad RDPUF, UDPUF: Reliable and unreliable parts of the DPUF
\State $t$: Confidence threshold
\State Split $X$ into train/test sets: $o$ \& $p$
\State $F_a \gets$ Fine-tune VGG16 using $o$
\State \textit{/* $F_a$: Training features */}
\State $F_b \gets$ Fine tune VGG16 using $p$
\State \textit{/* $F_b$: Testing features */}
\State Train Classifier using $F_a$
\State Test Classifier using $F_b$
\State $DPAN \gets$ Combined trained VGG16 \& Classifier
\State $t \gets$ Tune confidence threshold to zero false positives
\State $RDPUF, UDPUF \gets$ Characterization \& Analysis
\State \textbf{return} $DPAN, t, RDPUF, UDPUF$
\end{algorithmic}
\end{algorithm}

\begin{algorithm}
\caption{Reliability Analysis}
\label{alg:reliability_analysis}
\begin{algorithmic}[1]
\State \textbf{Input:} $C \in \{c_0, c_1, \ldots, c_n\}$: Challenge Data
\State \textbf{Output:} $SC$: Stable challenge
\State \textbf{Data:} $P$: Environmental parameters (Temp., Voltage)
\State $n$: Number of challenges
\State $r$: Number of repeat measurements
\State $t_e$: Error threshold
\State $l$: Size of responses in bits
\While{$length(SC) < l$}
    \For{$i \gets 0 \text{ to } n$}
        \For{$j \gets 0 \text{ to } P$}
            \For{$k \gets 0 \text{ to } r$}
                \State $R_{ijk} \gets$ PUF measurement for $C_i$
                \State $HW_{ij} \gets$ Add $R_{ijk}$ to Hamming Weight
            \EndFor
            \For{$\text{Cell } b \text{ in } HW_{ij}$}
                \If{$b \text{ in } SC$}
                    \If{$HW_{b}/r < t_e \text{ or } > 1 - t_e$}
                        \State $SC \gets$ Remove b from $SC$
                    \EndIf
                \ElsIf{$HW_{b}/r > t_e \text{ or } < 1 - t_e$}
                    \State $SC \gets$ Append b to $SC$
                \EndIf
            \EndFor
        \EndFor
    \EndFor
\EndWhile
\State \textbf{return} $SC$
\end{algorithmic}
\end{algorithm}

% \todo[inline]{The definitions and secure definitions, theorems or lemmas}
% \begin{definition}[PUF Phenotype]
%     Let $DPUF_{i}$ be a DRAM PUF, we say $Phenotype_{DPUF_{i}}$ is the PUF-Phenotype of $DPUF$ if it is generated in \textbf{Algorithm} \ref{alg:PUF-Phenotype}.
% \end{definition}

% \begin{definition}[DPAN]
%     Let $Phenotype_{DPUF_{i}}$ be the PUF-Phenotype of $DPUF$, we call $DPAN_i$ is the DRAM Phenotype-based Authentication Network (DPAN) if it is generated in \textbf{Algorithm} \ref{alg:DPAN}.
% \end{definition}

% \begin{theorem}[PUF Phenotype]
%     (Reveal no information about the PUFs' responses)
% \end{theorem}

% \begin{theorem}[DPAN]
%     (Classification accuracy, etc.)
% \end{theorem}

\subsection{Reliability Analysis}
For the proposed protocol, the data generated for DPAN training must also be analysed to locate a map of highly stable bits, which can be used to form noise-free responses using minimal error correction. The process for doing so is a modified version of that used in \cite{miskelly2020fast} and can be performed on the prover device as the DPAN training data is generated. The process for generating one stable response for a single device is described in \ref{alg:reliability_analysis}. The plain language description is as follows: Each response must be generated a large number of times and the Hamming weight for each cell's response recorded. This can then be used to locate bits above a certain threshold of stability. The process is repeated for each temperature and voltage point, retaining only bits which exceed the threshold in every case. In \cite{miskelly2020fast} a threshold of 0.99 (99\%) found at least 2\% of the total memory to be suitable. The final map of these high-stability cells produces a challenge which will result in a response of reliability no less than the threshold value across all operating conditions. The choice of threshold is a design consideration based on acceptable error rate and available resources. A higher threshold means the stable response will be spread out over a larger memory region and take longer to regenerate, but will have a lower bit error rate and therefore, require less correction.  

%As shown in Figure \ref{fig:usage_DPUF}, we test the DRAM PUF under different environmental effects, e.g. temperature and voltage.

\begin{figure}
    \centering\includegraphics[width=0.9\linewidth]{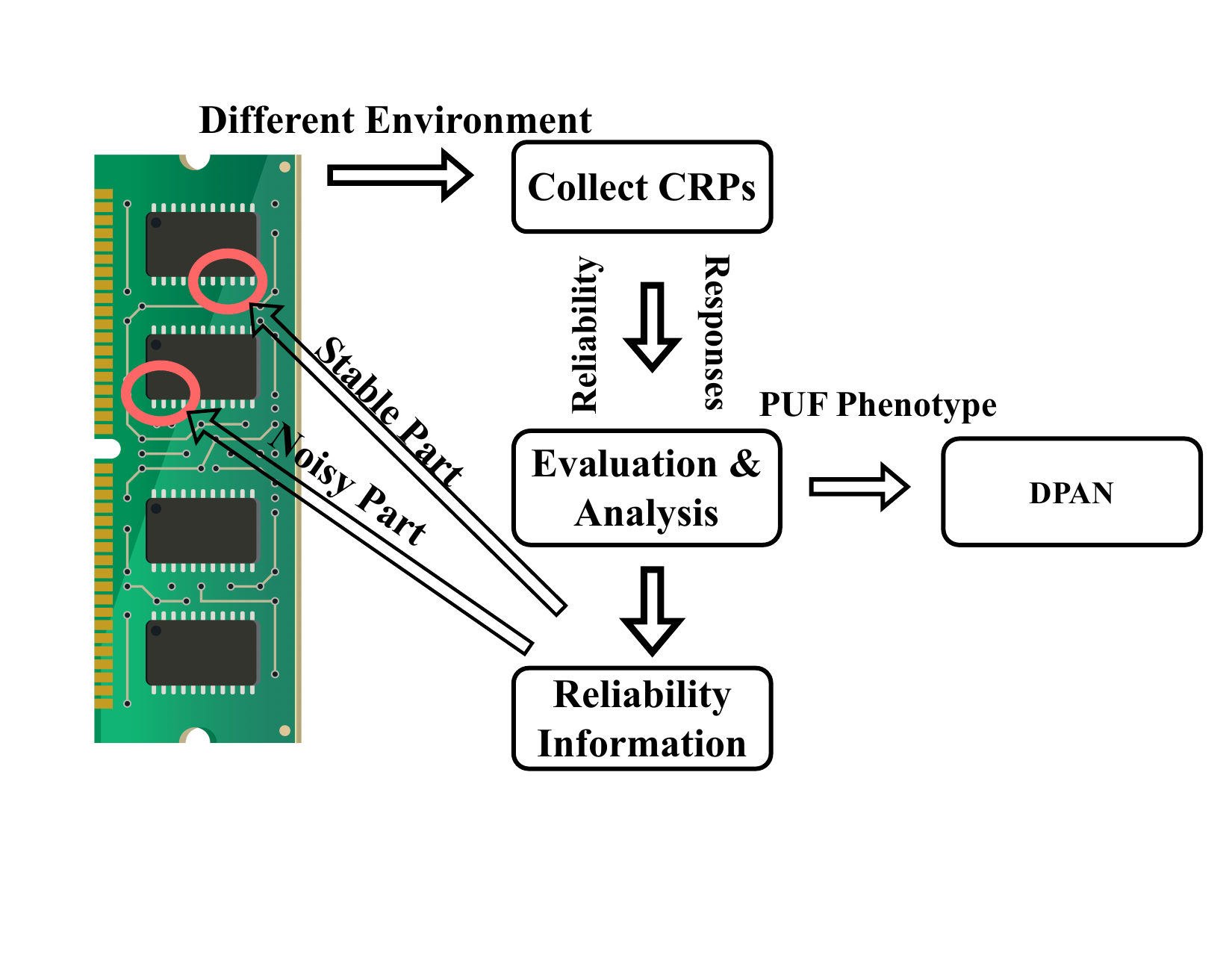}
    \caption{Usage of DRAM PUF in the proposed protocol.}
    \label{fig:usage_DPUF}
\end{figure}
\section{Proposed Scheme}
\label{sec:scheme}
In this section, we describe our proposed PUF-Phenotype-based authentication protocol.

\subsection{System Model}

Our system model encompasses a typical IoT communications scenario where a group of IoT devices can engage in direct interactions among themselves and establish communication with their gateway or vice versa. In this regard, any device/gateway can act as either a Prover or a Verifier. In such a group, each device stores information derived from other devices in the network during the enrollment phase. It should be noted that the stored information performs as the identity list ($Dev_{id}$) and enrollment information ($DPAN,\Delta$), which does not take much space. More performance discussion is presented in Section \ref{sec:performance}. During the authentication, devices will authenticate each other's identity and agree on an identical key for further communication using a combination of the stored data and noisy PUF responses generated as needed. At the end of each session, devices will update the interior states. Because of tamper resistance, DRAM PUFs cannot be modified or read out when the protocol is not executing, i.e., when the device is not powered on. We want the whole system to have the same security property of PUFs, tamper resistance, which means the ability to resist physical and logical attacks to alter, duplicate, or extract the sensitive information they contain. Besides, we consider a group authentication case, where a branch of devices communicate with each other. We want to ensure that even when a single device is corrupted after the enrollment phase, the security of the other devices will not be affected.

\begin{figure}
    \centering
    \includegraphics[width=0.9\linewidth]{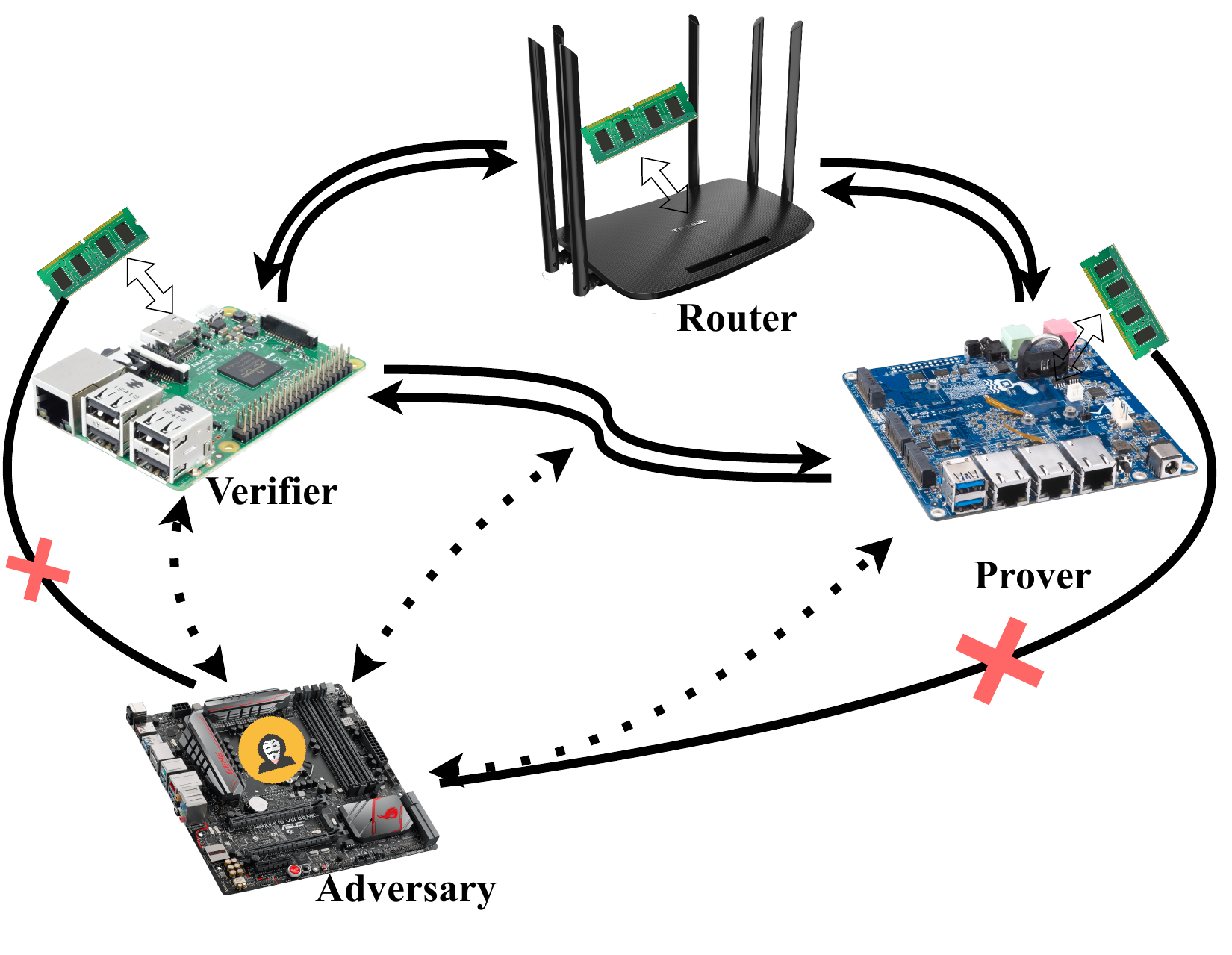}
    \caption{System model.}
    \label{fig:system_model}
\end{figure}

\subsection{Adversary Model and Assumptions}\label{AM_informal}
For the implemented IoT communication scenario, we consider two adversary models. First, we allow the adversary to have the capability described in the typical Dolev-Yao intruder model\cite{dolev1983security}, who has complete control over radio links in the network. The adversary can eavesdrop, alter, or block any messages he/she wants. The adversary has all the knowledge needed to understand the content in transferred messages. For example, the adversary knows what $M_1,M_2$ are composed of. We allow the adversary to have physical access to the device when the protocol is not executed, which means that the adversary can read, write or modify any information inside the non-volatile memory (NVM). After the enrollment and before the execution of the authentication protocol, the adversary can eavesdrop and alter any data he/she wants. However, we do not allow the adversary to physically access the DRAM PUF. Considering the property of tamper resistance, it is reasonable to assume that the adversary cannot perform such attacks directly on PUFs. Third, we also consider the case that, at some moments of authentication protocol execution, the adversary gets full control of one device, including accessing the DPUF arbitrarily, which means that he/she can obtain the key source.

We analyse the proposed protocol under the following assumptions.
 \begin{enumerate}
     \item First, the DPUF is assumed to be secure due to its tamper resistance property, which means any attempt to readout or modify the DPUF will be detected and will cause the whole authentication to fail.
     \item Second, the intermediate variables or states generated during the execution of the authentication protocol cannot be obtained, altered or erased.
     \item Third, we consider the adversary to have the ability to corrupt one device in the execution of the authentication protocol. We assume that all the challenges used in our protocol $C_{i}$ is only used once, which ensures that all the other devices can preserve all security properties apart from communicating with the corrupted one.
 \end{enumerate}
% We ensure the whole system has the same security property of PUFs, tamper resistance, which means the ability to resist physical and logical attacks aimed at altering, duplicating, or extracting the sensitive information they contain. 

% \begin{enumerate}
%     \item If considering insecure NVM, it cannot provide backward or forward security. If the adversary gets history $R_d \oplus R_s$, he can modify the information stored in the NVM, stopping the session key from updating. If the session key for one session is corrupted, then all the sessions, including history and future ones, are not secure anymore. \textbf{To address this flaw, data integrity verification should be added, e.g. MAC, AEAD.} An alternative way is to store the authentication information on both sides.
%     \item Against semi-honest devices, the adversary can learn the device's responses through XOR operations. Thus, it is necessary to generate different challenges for different devices for authentication. (Next we allow the adversary to participate in a honest execution in an honest execution of the proposed protocol, break one device and break the whole group of devices, remove the items, use next and afterward, step by step.)
% \end{enumerate}

\subsection{Proposed Authentication Protocol}
The proposed protocol consists of two phases: the enrollment phase and the authentication phase. The protocol has two participants, i.e., the prover and the verifier. In the enrollment phase, the prover and the verifier exchange responses on the same challenge, and then they store the device ID, XORed responses from both sides, and corresponding challenges in the NVM.

\begin{figure}[ht!]

\centering
\input{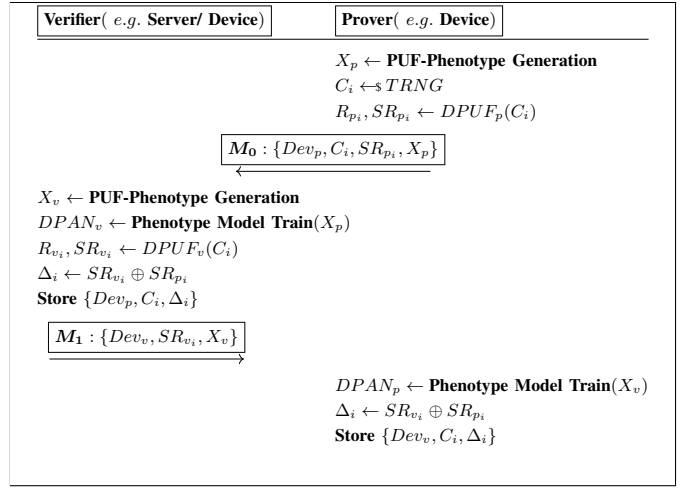}
\caption{The enrollment phase of the proposed protocol}
\label{fig:enroll}
\end{figure}

\subsubsection{Enrollment Phase} As shown in Fig. \ref{fig:enroll}, all the devices (may act as a prover or verifier) in the proposed IoT ecosystem need to exchange all the security parameters and other relevant information, such as model parameters (that will be used during the authentication phase) using the following three steps and through a secure communication channel. The adversary is not supposed to get any advantage in this phase.
\begin{itemize}
    \item \textbf{Step 1:} In this step, the prover initializes the enrollment. We refer to the session id as $i$. First, the prover generates the challenge $C_i$ using a true random number generator (TRNG), forwards it to the DRAM PUF, and gets the noisy response $R_{p_i}$ and stable response $SR_{p_i}$. Then, it invokes the PUF-Phenotype Generation algorithm to generate the Phenotype dataset. The prover sends the combined information $M_0$, including the device id $Dev_{p}$, challenge $C_i$, the stable response $SR_{p_{i}}$ and Phenotype dataset $X_p$, to the verifier.
    % \todo[inline]{The DPAN enrollment phases and model training need more description.}
    \item \textbf{Step 2:} In this step, the verifier generates a session key source and PUF Phenotype dataset for further pairwise authentication. After receiving the request from the prover, the verifier invokes the DPAN Enrollment Phase and generates the Phenotype dataset $X_v$. Secondly, it invokes the Phenotype model training algorithm to get the trained model $DPAN_v$ from $X_p$. Then, the verifier inputs the challenges $C_i$ to the $DPUF_{v}$ and gets the noisy response $R_{v_i}$ and stable response $SR_{v_i}$. By XORing stable responses from both side, the session $\Delta_i$ can be generated from $SR_{v_i}$ and $SR_{p_i}$, which can be stored directly without leaking any credential information. The verifier also stores the prover's ID $Dev_{p}$ for later checking. In the end, the prover sends its id $Dev_{v}$, static response $SR_{v_i}$, and Phenotype $X_v$ back to the prover. It should be noted that, for enrollmetn of a group of devices, the model training process needs all the PUF-Phenotype from all the engaging devices' DPUFs. Also, it does no harm on the security of the protocol even if the adversary can get the contents stored in the NVM.
    \item \textbf{Step 3:} After receiving $M_1$ from the verifier, the prover performs similar operations. First, it trains the model by invoking the Phenotype model training algorithm and gets $DPAN_p$. Then it computes XORed stable responses from both sides and stores it. It also stores the verifier's id together with the challenge $C_i$. 
\end{itemize}

\subsubsection{Authentication Phase:} 
\begin{figure}[ht!]
\centering
\input{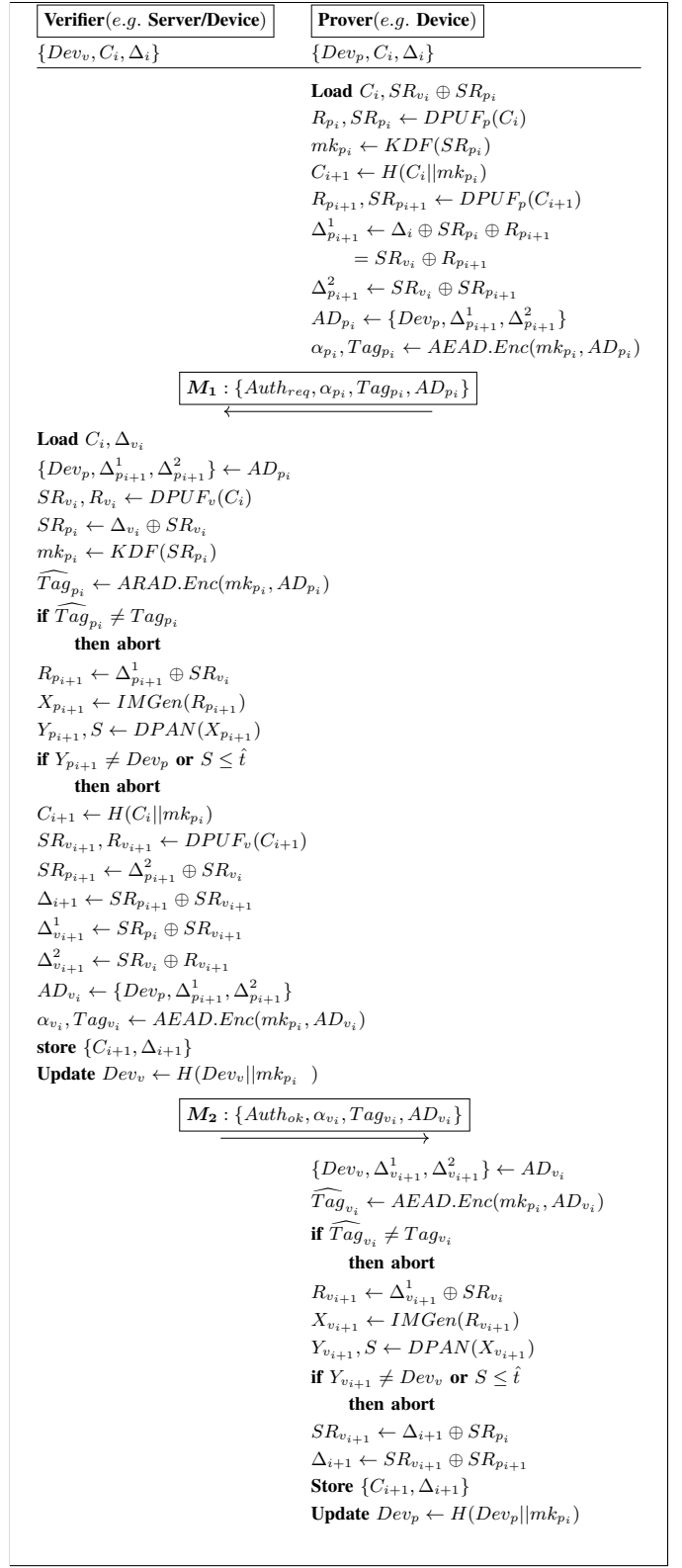}
\caption{Authentication phase of proposed protocol.}
\label{fig:auth_new2}
\end{figure}

As shown in Fig. \ref{fig:auth_new2}, the prover and verifier communicate with each other in 3 steps and then authenticate each other.

% \begin{figure}[ht]
% \label{fig:auth}
% \centering
% \input{figures/authentication}
% \caption{Authentication Phase of Proposed Protocol}
% \end{figure}

% \begin{figure}[ht]
% \label{fig:auth_new}
% \centering
% \input{figures/authentication_new}
% \caption{Authentication Phase of New Proposed Protocol}
% \end{figure}

% \begin{table}[ht]
%     \caption{Symbols and Descriptions}
%     \label{tab:symbols}
%     \centering
%     \begin{tabular}{cc}
%     \toprule
%          \textbf{Symbol} & \textbf{Description} \\
%     \midrule
%          &  \\
%     \bottomrule
%     \end{tabular}
% \end{table}

\begin{itemize}
    \item \textbf{Step 1:} In this step, a device that wants to authenticate with a target device $Dev_v$ initiates the protocol. In this session, the initializer device will play the role of a prover, and the other be the verifier. The prover first loads stored information from the enrollment phase, $C_i,SR_{v_{i}},\Delta_i$. Then it inputs the challenge $C_i$ to the DPUF and gets two parts of responses, the noisy response $R_{p_i}$ and stable response $SR_{p_i}$. $SR_{p_i}$ is viewed as the entropy source and used to derive the session key $mk_{p_i}$. Then, it computes the new challenge for the next session as $C_{i+1}\gets H(C_{i})||mk_{p_i}$. The adversary cannot trace the value of the challenge if the session key is secure. Then, the prover inputs $C_{i+1}$ to the DPUF and gets new noisy response $R_{p_{i+1}}$ and stable response $SR_{p_{i+1}}$, and calculates $\Delta_{p_{i+1}}^1$ and $\Delta_{p_{i+1}}^2$ by XORing the current stable response with the new noisy response, and current stable responses with the new stable response, respectively. The prover packs the device id $Dev_{p_i}$, $\Delta_{p_{i+1}}^1$ and $\Delta_{p_{i+1}}^2$ as the associated data, and encrypts them with session key $mk_{p_i}$. It can get the ciphertext $\alpha_{p_i}$ and authentication Tag $Tag_{p_i}$. In the end, the prover sends $M_1$ to the verifier, including a request for authentication, $\alpha_{p_i}$, $Tag_{p_i}$ and $AD_{p_i}$, and waits for the response.
    
    \item \textbf{Step 2:} In this step, the verifier decides whether or not to accept the authentication; if so, it will generate the PUF Phenotype and send its identity information to the device for authentication. After receiving $M_1$, the verifier loads $C_{i},\Delta_{i}$ according to the ID $Dev_{p_i}$. If it exists, it will unpack the received associated data and get $\alpha_{p_i}$, $Tag_{p_i}$ and $AD_{p_i}$. The first thing the verifier does is to verify whether $M_1$ is honestly transferred. It inputs $C_i$ to its DPUF $DPUF_{v}$ and gets noisy response $R_{v_i}$ and stable response $SR_{v_i}$. Then, it can get $SR_{vp_i}$ by XORing $\Delta_{i}$ and $SR_{v_i}$. The session key can be derived from $SR_{v_i}$ using \textit{Key Derivation Function} (KDF). The tag can be verified using the session key. The verifier aborts the protocol if $\widehat{Tag_{p_i}}\neq Tag_{p_i}$, which means $M_1$ has been modified or is unsynchronized. If verification passes, the verifier will test whether this $M_1$ is generated by an honest prover. It first unpacks the noisy response $R_{p_{i+1}}$ by XORing $SR_{v_i}$ and $\Delta_{p_{i+1}}^1$. Secondly, it invokes $IMGen(\cdot)$ to generate an image from the noisy response, which will be the input of the $DPAN$ model. The $DPAN$ model outputs the classification result $Y_{p_{i+1}}$ and the confidence value $S$. The protocol will be aborted if the classification result is wrong or $S$ is lower than the threshold $\hat{t}$. If the verification passes, the verifier will believe that the message $M_1$ is honestly transferred and sent by a legal prover. Here, the verifier can obtain the updated stable response of the prover by $SR_{p_{i+1}}\gets \Delta_{p_{i+1}^2} \oplus SR_{v_i}$. Then it first generates the new challenge $C_{i+1}\gets H(C_i||mk_{p_i})$, where $H(\cdot)$ is a hash function. Then the prover generates the new noisy response $R_{p_{i+1}}$ and stable response $SR_{p_{i+1}}$. It can update $\Delta_{i+1}\gets SR_{p_{i+1}}\oplus SR_{v_{i+1}}$. $\Delta_{p_{i+1}}^1$ and $\Delta_{p_{i+1}}^2$ can be generated from $SR_{p_{i}}\oplus SR_{v_{i+1}}$ and $SR_{v_{i}}\oplus R_{v_{i+1}}$. Then, the verifier packs $Dev_{v},\Delta_{p_{i+1}}^1$ and $\Delta_{p_{i+1}}^2$ as the associated data $AD_{v_i}$, encrypts them with the session key $mk_{p_{i}}$, and gets the ciphertext $\alpha_{v_i}$ and authentication tag $Tag_{v_i}$. In the end, the verifier stores $C_{i+1},\Delta_{i+1}$, and updates the id $Dev_{v}\gets H(Dev_{v}||mk_{p_i})$. $M_2$ is then sent, composed of the response of the authentication request $Auth_{ok},\alpha_{v_i},Tag_{v_i}$ and $AD_{v_i}$. It should be noted that, as long as the session key is not compromised, the leakage of $\Delta,\Delta^1,\Delta^2$ does no harm to the security of the protocol. A formal secuirty proof is presented in Section \ref{sec:sec_ana}.
    
    \item \textbf{Step 3:} In this step, the prover will validate two things, the first is whether $M_2$ is honestly transferred, and the second is whether $M_2$ is generated in this session by a legal verifier. After receiving $M_2$, the prover first unpacks the associated data into $Dev_{v},\Delta_{v_{i+1}}^1,\Delta_{v_{i+1}}^2$. $AD_{v_{i}}$ can be verified by encrypting it with the session key $mk_{p_i}$, and then checking if the generated tag $\widehat{Tag_{v_i}}$ equals $Tag_{v_i}$. Secondly, the verifier obtains the noisy response of the verifier by $R_{v_{i+1}}\gets \Delta_{v_{i+1}}\oplus SR_{v_{i}}$. Similiar with the verifier, the prover invokes $IMGen(\cdot)$ and $DPAN$ model to verify the identity of the verifier. The session will be aborted if either the classification result is wrong or the confidence value is below the threshold. If they all pass, the prover updates its stored data. The new $\Delta_{i+1}$ for the next session can be generated by XORing $SR_{v_{i+1}}$ and $SR_{p_i+1}$. The device ID can be generated by $Dev_{p}\gets H(Dev_p||mk_{p_i})$. The updating of the device id can ensure forward and backward privacy as long as the assumption of this protocol stands. At this point, all the steps on the device side have been completed.
\end{itemize}

\begin{figure}
\centering
\begin{adjustbox}{max width=0.50\textwidth, max height=0.5\textheight}
\input{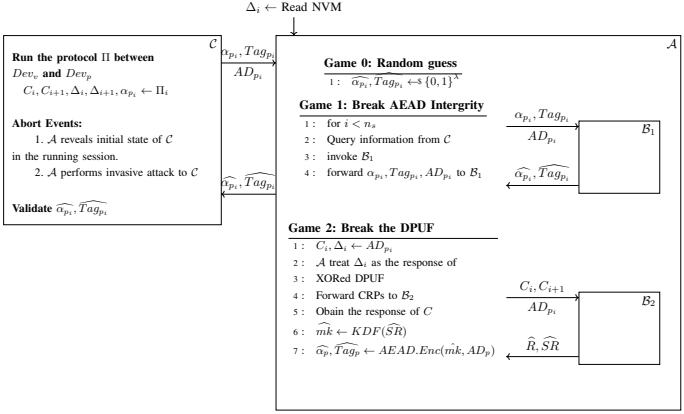}
\end{adjustbox}
\caption{Security Framework for Mutual Authentication.}
\label{game:MU}
\end{figure}

\section{Security Model and Analysis}
\label{sec:sec_ana}
\subsection{Adversarial Model}
\label{adversary model}
Based on the discussion in Section \ref{AM_informal}, we formally define the invasive adversary to analyse the security of the proposed protocol. We allow the adversary to eavesdrop, modify, and delay communication between the prover and the verified to gain the advantage of breaking the security. The adversary can also perform invasive operations on the on-chip NVM, which means it can readout the stored information in the NVM. It is also allowed to modify the data. Also, if the adversary blocks the entire communication channel (which can be detected easily), the authentication will fail, and the best result for the adversary is the same as the denial of service (DoS) attack. Since we assume that the enrollment phase is performed in a secure channel, we consider $\adv$ can issue the following Oracle queries:
\begin{enumerate}[label=$\cdot$]
    \item \textbf{Launch}$(1^{\lambda})$: A new session is started by the prover.
    \item \textbf{Send}$\prover(Dev_{id},m)$: Send arbitrary message $m$ to device $Dev_{id}$.
    \item $\text{\textbf{Issue}}(DPUF)$: Allows $\adv$ to have physical access to the PUF instance $DPUF$, and perform queries \textbf{Challenge}$(\cdot)$ and \textbf{Reveal}$(\cdot)$.
    \item $\text{\textbf{Reveal}}(C_{i}, \Delta_i)$: $\adv$ reveals sensitive data $C_{i}$ and $\Delta$ in NVM.
    \item $\text{\textbf{Corrupt}}(\cdot)$: Allows $\adv$ to modify the content in NVM.
    \item \textbf{Block}$(\cdot)$ Allows $\adv$ to block authentication massage $M_1,M_2$ for any side it wants in running of the protocol.
\end{enumerate}
We allow $\adv$ to have access to the NVM only when the session is not running on the device, and forbid it from accessing the DPUF. Thus, we give the definition for a \textbf{cleanness predicate}.
\begin{definition}[Cleanness predicate]
    If the $\adv$ does not invoke \textbf{Issue}$(puf)$ during the whole game, and only invokes \textbf{Reveal}$(\cdot)$ and \textbf{Corrupt}$(\cdot)$ when the protocol is not running on the device, we call this session a \textbf{cleanness predicate}.
\end{definition}

Apart from the cleanness predicate defined above, we also state which kind of interactions between adversary $\adv$ and challenger $\mathcal{C}$ are \textbf{clean}, in order to capture the nontrivial advantage $\adv$ can obtain without breaking the security of our protocol. Next, we give the definition of a matching session.

\begin{definition} [Matching Session]
    In the device-to-device (D2D) communication protocol between a prover $Dev_{p}$ and a verifier $Dev_{v}$, we say a session $\pi_{pv_i}$ is a matching session if device id $Dev_{p}$ and $Dev_{v}$ are valid and the messages are honestly transferred between these two valid devices.
\end{definition}

\subsection{Security Model}
\label{sub:security_model}
Here we describe the security analysis model for the protocol. In the model, the challenger $\mathcal{C}$ maintains a group of IoT devices $D_{1},\cdots,D_{n}$, with each device operating a number of instances of the proposed mutual authentication protocol $\Pi$. We denote the protocol performed between device $p$ and device $v$ as $\Pi_{pv}$, and the $i$-th session as $\pi_{pv_{i}}$. We define a matching session where the messages communicated between devices are honestly transferred. We capture the protocol's security $\Pi_{pv}$ through a game performed between a challenger $\mathcal{C}$ and an adversary $\adv$, noted as $\mathbf{Exp}_{\Pi,\adv}^{MA}(\lambda)$. The goal of $\adv$ is to model the communicated message $M_{1}:\{\alpha_{p_{i}},Tag_{p_{i}},AD_{p_i}\}$ and $M_{2}:\{\alpha_{v_{i}},Tag_{v_{i}},AD_{v_i}\}$ or derive the session key $mk$ used in $\Pi_{pv}$. At the end of the authentication process, $\mathbf{Exp}_{\Pi,\adv}^{MA}(\lambda)$ will output 1 (accept) or 0 (reject). We say $\adv$ wins the game if it causes a \textbf{clean} session and $\mathbf{Exp}_{\Pi,\adv}^{MA}(\lambda)$ output 1, where $\mathcal{C}$ accepts the authentication without a matching session.  The basic game is considered as follows:

$\mathbf{Exp}_{\mathbf{\Pi,A}}^{MU}(\lambda)$:
\begin{enumerate}
    \item $(Dev_{id},C_{i}, \Delta_{i},pp)\xleftarrow[]{Random}\text{Setup}(\cdot)$;
    \item $(\hat{Dev_{id}},\widehat{M}_{1},\widehat{M}_{2},\widehat{mk_{i}})\xleftarrow[]{Random} \adv^{\text{Issue, Challenge, Reveal, Corrupt,Send$\prover$,Block}}((C_{i}, \Delta_{i},pp))$;
    % \item $S^',hd^'\xleftarrow[]{Random} \text{Reconfig}(\cdot)$
    \item $\Phi := Outcome(\widehat{mk_{i}},\widehat{M}_{1},\widehat{M}_{2})$;
\end{enumerate}

The challenger $\mathcal{C}$ executes a setup algorithm for enrolling into a trusted environment and all the public parameters $pp$ for initialization. Here, $pp$ denotes all the available public parameters used to initialize the authentication, e.g., the length of security parameters and the setting up of KDF. The basic game of mutual authentication comprises of two parts: prover authentication and verifier authentication. For both parts, we consider the same adversary model as discussed in Section \ref{adversary model}. We denote the advantage of the adversary $\adv$ in this game as $\advantage{MU}{\Pi,\adv}[(\lambda)]$, which is defined as the probability $\adv$ causes a winning event.

\subsection{Privacy Model}
\label{app:privacy}
\begin{figure}
\centering
\begin{adjustbox}{max width=0.49\textwidth, max height=0.49\textheight}
\input{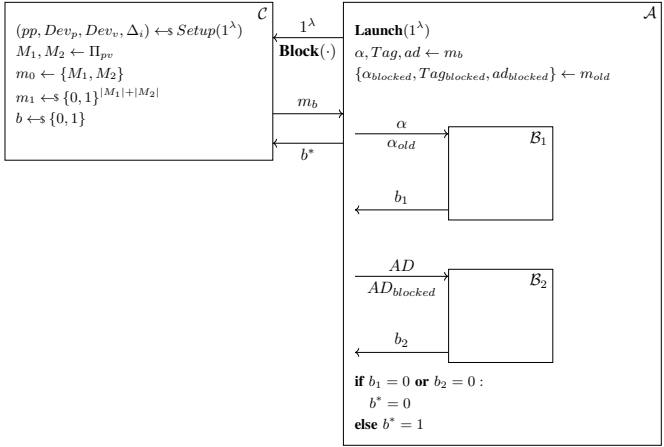}
\end{adjustbox}
\caption{Privacy game.}
\label{game:privacy}
\end{figure}

Now, we consider indistinguishability-based privacy. In this case, the challenger $\mathcal{C}$ maintains 2 devices that communicate with each other using the proposed protocol. The adversary $\adv$ eavesdrops somewhere in the network but tries to trace the communication parties. The privacy goal is that, in a clean session, the adversary $\adv$ cannot distinguish the true message from a randomly generated value. The basic game is shown in Figure \ref{game:privacy} and considered as follows:

$\mathbf{Exp}_{\mathbf{\Pi,A}}^{\mathrm{IND}}(\lambda)$:
\begin{enumerate}
    \item $\adv$ issues \textbf{Launch}$(1^\lambda)$.
    \item $\mathcal{C}$ sets up the system, starts running the protocol $\Pi_{pv}$ on devices $p$ and $v$, and obtains $M_1,M_2$ from the communication of the protocol.
    \item $\mathcal{C}$ combines $M1,M2$ as $m_0$, and then generates a random value with the same length $|M_1|+|M2|$, noted as $m_1$. Then, $\mathcal{C}$ generates a random choice $b$ from $\bin$, and then sends $m_b$ to $\adv$.
    \item $\adv$ sends his choice $b^*$ to $\mathcal{C}$.
\end{enumerate}
At the end of the game, $\mathcal{\mathbf{Exp}_{\mathbf{\Pi,\adv}}^{\mathrm{IND}}(\lambda)}$ outputs the result whether $b=b^*$, 1 if the same and 0 if not. We say $\adv$ wins the game if $\mathbf{Exp}_{\mathbf{\Pi,\adv}}^{\mathrm{IND}}(\lambda)$ outputs 1, and note that the advantage that $\adv$ wins the game is $\advantage{\mathrm{IND}}{\Pi,\adv}=|\prob{(\mathbf{Exp}_{\mathbf{\Pi,\adv}}^{\mathrm{IND}}(\lambda)=1)}-\frac{1}{2}|$.

\subsection{Mutual Authentications}
We divide the analysis into two cases: prover impersonation and verifier impersonation. In prover impersonation, the protocol $\Pi$ accepts $M_1$ without an honest matching partner. We capture the security between a challenger $\mathcal{C}$ and adversary $\adv$. Figure \ref{game:MU} shows the whole game $\mathbf{Exp}_{\mathbf{puf,A}}^{\mathrm{MU}}(\lambda)$. Here, we show the analysis step by step, taking prover impersonation as an example.

\textbf{Prover Impersonation:}
Now, we consider the above model for analyzing the security of the proposed authentication protocol.

\textit{Proof:}  As discussed in Section \ref{sub:security_model} and shown in Figure \ref{game:MU}, the adversary $\adv$ tries to mimic an honest prover by generating  $mk$ and $M_1$, without a corresponding matching session and also in a cleanness predicate. We start from the original game.

\textbf{MUGame 0:} This is the original game played between $\mathcal{C}$ and $\adv$. We denote the advantage of this game as $\advantage{MUGame0}{\Pi,\adv}$.

\textbf{MUGame 1:} In this game, $\adv$ replaces $mk_{p}\sample \bin^{|mk_{p}|}$. First, $\adv$ physically assesses a device before the session starts and reads $Dev_{p},C_{i},\Delta_{i}$ from the NVM. Then it issues \textbf{Launch}$(1^\lambda)$ and tries to generate $M_1$. It should be noted that $\adv$ cannot access the NVM when the device is running. During the communication, $\adv$ eavesdrops on all the communicated information and stores it. After it collects a polynomial amount of data, it blocks the device's communication and tries to mimic it. For thet, the adversary collects associated data $\{AD\}$, the tags $\{Tag\}$ and the ciphertext $\alpha$, and invokes $\bdv_{1}$, which is built to break the security of AEAD algorithm. In the proposed protocol, the session key is updated after every session, and $\adv$ can obtain no advantage from knowing the old key for a new session; even $\bdv_1$ breaks the CPA-security of AEAD. The only chance is that $\bdv_1$ can produce ciphertext and tag without knowing the correct key, which is bounded by the integrity game. Here, we introduce an abort event when $AD_{p_i}$ is inconsistent with the tag $Tag$. Thus, the advantage after this game is:
\begin{equation}
    \advantage{MUGame1}{\Pi,\adv}<\advantage{MUGame0}{\Pi,\adv}+\epsilon_{1}+\advantage{AEADauth}{\bdv,\bdv_1} .
\end{equation}

\textbf{MUGame 2:} In this game, $\adv$ treats $\Delta_{i}$ as the XORed responses from two DPUFs. The adversary has already collected enough CRPs from the associated data in \textbf{Game 1} and it issues the adversary $\bdv_{2}$ and forwards the CRPs to it. Then, $\adv$ gets the output $\widehat{R},\widehat{SR}$ from $\bdv_{2}$. $\adv$ invokes $KDF$ to derive the session key $\widehat{mk}$ from $\widehat{SR}$. Finally, as $\adv$ has the `session key' and associated data $AD_p$, it can generate $\widehat{\alpha_{p}},\widehat{Tag_p}$ using $AEAD.Enc(\cdot)$. The advantage of this game is:
\begin{equation}
    \advantage{MUGame2}{\Pi,\adv}\leq\advantage{MUGame1}{\Pi,\adv}+\advantage{ModPUF}{\Pi,\adv} .
\end{equation}

We can find that if the session key is secure, then $\adv$ cannot break the authentication security in the verifier impersonation game if the $AEAD$ algorithm is secure, and $DPUF$ is secure against the modelling attack.

\textbf{Verifier Impersonation}
Now, we analyse the security of the proposed authentication protocol from the view of a prover. The goal is that the adversary $\adv$ cannot start a \textbf{clean} session and win the $\mathrm{MU}$ without a matching session.\\
\textit{Proof:} Similar to the discussion in the prover impersonation case, the adversary $\adv$ tries to mimic an honest verifier by generating the session key $mk$ and $M_2$. In the proposed protocol, the prover and verifier store the same information on the device, and the adversary is allowed to have physical access to the NVM when the protocol is not running. The transferred information $M_1$ and $M_2$ are highly similar, with the exception of the authentication state ($Auth_{req},Auth_{ok}$) and associated information $AD$. Thus, the analysis is similar, and the challenger $\mathcal{C}$ will abort the game when the tag cannot be verified or the protocol aborts itself as discussed in the protocol.
We present the full formal analysis in the Appendix\footnote{\href{https://drive.google.com/file/d/1FOmiMai0W5AjRuvvo2bVQUnXPsBHLODF/view?usp=drive_link}{\textbf{Link for the Supplementary Material (Complete Security Proof)}}} for privacy security. An informal discussion is given in Section \ref{sub:sec_eval}.

\section{PERFORMANCE ANALYSIS AND COMPARISON}
\label{sec:performance}

% \todo[inline]{Notes for this section. Drawback: This needs a lot of resources over older PUF protocols - hash functions, KDF, etc. Counterpoints: PUF protocols that aren't trivial to break often also need these. Benefit: This allows the system to use all of the memory with minimal error correction and even make use of the cells, which are almost entirely random. Contrast memory PUF normally where only the stable response part can be used this way, and up to 98\% is noisy. Is that gain worth the cost of the scheme? Does that property provide a meaningful security advantage? How does this compare to non-PUF key agreement and authentication schemes? What benefit does it add? Benefit: DPAN model plus deltas don't take up too much memory; we're talking tens of MB, which makes this a viable device-to-device scheme. How does it compare against other D2D schemes? Benefit: The group-based DPAN model is the same on every device and leaks no info if compromised. Counterpoint: The need for device-specific stable responses on each device weakens this slightly. This means that to add a new device, you must train the model repeatedly and deploy it to all devices. Counter-counterpoint: Retraining is trivial if you keep the enrollment data somewhere central. Something to note: The DPAN model for larger numbers of CRPs has not been experimentally proven, so we're assuming it scales up. Will it be in practice? It remains to be seen.}

\subsection{Security Evaluations}
\label{sub:sec_eval}
This section compares the proposed protocol with five state-of-the-art PUF-based authentication protocols. It can be observed that most of these protocols are based on assumptions that may not stand in practice, e.g., some protocols demand a trusted third-party verifier, a secure NVM, etc. Here, we have listed several vital properties ($P_1$-$P_{10}$) expected to build a secure IoT ecosystem. Next, we use these properties ($P_1$-$P_{10}$)) to benchmark the performance of our proposed scheme with others, as shown in Table \ref{tab:comparison}. Next, we briefly explain how the proposed protocol achieves all the properties $P_1$-$P_{10}$). 

$P_1$ \textit{(Privacy and Confidentiality)}: In the IoT ecosystem, IoT devices often collect sensitive data (e.g., health, environmental etc.), and privacy is essential to maintain user trust and comply with data protection regulations. On the other hand, in some cases, the privacy of the devices is also vital, especially when we consider passive attackers. In our proposed protocol, the prover and verifier update their ID after every session $Dev_{id_{i+1}}\gets H(Dev_{id_{i}}||mk_{i})$. Hence, it will be difficult for a passive adversary to target a specific device. Besides, our proposed protocol ensures a secure session key establishment that helps to achieve the confidentiality of the messages transferred between the prover and verifier and vice versa.  On the contrary, in several state-of-the-art schemes, such as Zheng et al. \cite{Zheng2022P2P}, the IDs for devices $A$ and $B$ are sent in plaintext in the air; therefore, the adversary can easily trace the identity of communication parties.

$P_2$ \textit{(Mutual Authentication)}: This property is essential to establish trust between the prover and the verifier. In our proposed scheme, the verifier authenticates the prover by verifying the authentication tag $Tag_{p_i}$. Similarly, the prover authenticates the verifier using the $Tag_{v_i}$ authentication tag.

%In our proposed protocol, we can see from $M_1$ and $M_2$ that two mechanisms secure the authentication. $AEAD$ is used to guarantee the messages are honestly transferred. $DPAN$ is used to verify the source of the message in case the message is somehow forged or replayed. As long as the session key $mk$ is secure and the DPUFs of the devices can not be assessed by the adversary, we can achieve mutual authentication. The formal analysis is provided in Section \ref{sec:sec_ana}.

$P_3$ and $P_4$ \textit{(Backward Security and Forward Security)}:  Both these properties ensure the security of the whole protocol, even if the secrets ($mk$ for our proposed protocol) maintained by the entities are compromised by the adversary. In our protocol, $mk_p$ is derived from the prover's stable PUF response $SR_{p}$. As we limit the adversary's access to the DPUF, even if the adversary can learn the value of challenge, which can be obtained from the NVM or computed with $C_{i+1}\gets H(C_{i}||mk_{p_i})$, it still cannot generate the legitimate response of an unaccessible DPUF.  Thus, the compromise of the secret $mk$ will not help the adversary to decipher any previously or subsequent intercepted message. In this way, we achieve forward and backward security.

$P_5$ \textit{(No CRP Database Required)}: This property assesses the necessity of a CRP database. Studies by Yildiz et al. \cite{PLGAKD}, Gope et al. \cite{Gope2018RFID}, Zheng et al. \cite{Zheng2022P2P}, and Ren et al.\cite{Ren2021Group} all require IoT devices to store CRPs, pre-generated during the enrollment phase, in secure NVM. Consequently, the number of supportable sessions depends on the quantity of pre-shared and stored CRPs. In Yildiz et al. \cite{PLGAKD}, strong PUFs generate key entropy sources, claiming a vast key space. However, generating and storing CRPs can burden IoT devices (the server maintains the database in \cite{PLGAKD}). Our protocol, in contrast, stores the trained model $DPAN$ instead of CRPs, which can hardly be used to derive any CRP information. In this way, we mitigate risks associated with resource depletion and CRP leakage.

\begin{table}
\caption{Comparison of PUF Protocols}
\label{tab:comparison}
\centering
\begin{threeparttable}
\begin{tabular}{p{2.2cm} p{0.2cm}p{0.2cm}p{0.2cm}p{0.2cm}p{0.2cm}p{0.2cm}p{0.2cm}p{0.2cm}p{0.2cm}p{0.2cm}}
    \textbf{Schemes}
    & $P_{1}$
    & $P_{2}$
    & $P_{3}$
    & $P_{4}$
    & $P_{5}$
    & $P_{6}$
    & $P_{7}$
    & $P_{8}$
    & $P_{9}$
    & $P_{10}$\\
    \midrule
    \textbf{Yıldız et al.} \cite{PLGAKD} & \checkmark & \checkmark & \checkmark & \checkmark & x & \checkmark & x & x & \checkmark & x\\
    \textbf{Aysu et al.} \cite{AysuEnd2End} & \checkmark & \checkmark & \checkmark & \checkmark & \checkmark & x & \checkmark & x & x& x\\
    \textbf{Gope et al.} \cite{Gope2018RFID} & \checkmark & \checkmark & \checkmark & \checkmark & x & \checkmark & x & - & x& x\\ 
    \textbf{Zheng et al.} \cite{Zheng2022P2P} & x & \checkmark & x & \checkmark & x & x & x & x & \checkmark& x\\
    \textbf{Ren et al.} \cite{Ren2021Group} & \checkmark & \checkmark & x  & \checkmark & x &x &x & - &x& x\\
    \textbf{Yu et al.} \cite{yu2016lockdown} & x & \checkmark & x & \checkmark & x & x & x & \checkmark & x & x\\
    \textbf{Proposed Scheme} & \checkmark & \checkmark & \checkmark & \checkmark & \checkmark & \checkmark & \checkmark & \checkmark & \checkmark & \checkmark\\
    \bottomrule
    \end{tabular}
    \vspace{2pt}
    \begin{tablenotes}
    \footnotesize
    \item
    \checkmark: Yes; X: No; 
    \item
    $P_{1}$: \textbf{Privacy} $P_{2}$: \textbf{Mutual Authentication} $P_{3}$: \textbf{Backward Security} $P_{4}$: \textbf{Forward Security} $P_{5}$: \textbf{No CRP Database Required} $P_{6}$: \textbf{No Third Party Required} $P_{7}$: \textbf{No Secure NVM Required} $P_{8}$: \textbf{Error Correction without Using HD} $P_{9}$:\textbf{Supporting Group-based Authentication} $P_{10}$:\textbf{Group-based Authentication without Group Key}
    \end{tablenotes}
\end{threeparttable}

\end{table}
$P_6$ \textit{(No Third Party Required)}: This property relates to operating the protocol without a trustworthy third party, enhancing the autonomy of IoT systems and lessening their dependency on external networks or services, which is crucial for reliability and performance. D2D communication scenarios are more applicable in IoT applications. Our protocol can be highly symmetrical; any enrolled device can perform as the prover or verifier, realizing D2D communication.

Property $P_7$ \textit{(No Secure NVM Required)}: Since many IoT devices are deployed in the open and public places, it makes them vulnerable to invasive attacks. Therefore, any assumption on secure NVM is impractical. A sophisticated adversary could potentially read or modify NVM content, as highlighted in the CK-adversary model \cite{canetti2001analysis}. Our protocol ensures security and privacy, even with insecure NVM, as discussed in Section \ref{sec:sec_ana}. Unlike existing PUF-based protocols, in the proposed protocol, we allow the adversary to read the NVM when the protocol is not in execution. All it can obtain is the $Dev_{id},C_{i},\Delta_i$, which can neither be traced nor analysed since the ID changes after every session and the $C_{i},\Delta_i$ are useful only when DPUF can be assessed. 

Property $P_8$ \textit{(Error Correction without Using HD)}: Error correction is a common concern regarding authentication and key generation protocols which use PUFs as the entropy source. PUFs are sensitive to environmental effects, e.g., temperature and voltage, which means different responses can be observed at different system operating times. To deal with errors, the existing protocols rely on conventional error correction schemes (such as fuzzy extractor, majority voting, etc.). In contrast, our proposed scheme uses the DPAN model to deal with errors. 

$P_{9}$ and $P_{10}$  \textit{(Supporting Group Authentication without Group Key)}: In many IoT environments, such as home IoT or industrial IoT, sometimes a group of devices must communicate with each other securely. In this regard, considering the resource limitation of IoT devices, we need an efficient group authentication scheme. Unfortunately, existing PUF-based group authentication protocols rely on a group key or store multiple models on each device's memory. The protocols that rely on the group key are always vulnerable when the group key is compromised \cite{PLGAKD,Zheng2022P2P}. On the other hand, if we want to transform an 
 existing PUF-model-based approach \cite{yu2016lockdown} to support a group-based authentication system. In this regard, contemplating a
set of $n$ provers in a group, a designated verifier must retain the $n - 1$ model (or CRP database) to manage authentication requests from every other member of the group. This is inefficient as most of the IoT devices have limited storage and computational resources. In contrast, the proposed scheme efficiently supports group authentication, requiring minimal additional resources per device. Beyond the DPAN model, it necessitates only a single query of $Dev_{id},C,\Delta$ for each device. This demonstrates the proposed protocol's efficiency in terms of operational time and storage costs, particularly in the context of group authentication.

\subsection{Computational Cost}
In this section, we show the computation cost of the proposed scheme. In this regard, we enumerate all the cryptographic primitives utilized in our proposed protocol, along with their practical construction methodologies. Our protocol incorporates several components, namely: $\text{DRAM PUF}$, a \textit{Hash function}, an \textit{AEAD algorithm}, an \textit{Image generation algorithm}, a \textit{DPAN model}, and a \textit{Key Derivation Function}. These elements are specifically chosen for their minimal resource consumption, making them suitable for IoT devices. The computational cost of the protocol has been computed by aggregating the operational time of each function multiplied by its respective time cost. We denote the runtime of a function \(f(\cdot) \) as \( N_{f(\cdot)} \). Hence, the total estimated operational time can be expressed as shown in Table \ref{tab:cost_pro}. Given that our protocol is designed for D2D communication scenarios, the time expenditure for both parties involved is approximately equal to the calculation above. To rigorously analyse the proposed protocol's performance, particularly the device overhead associated with running the DPAN model needs to be evaluated. For this, we use similar implementation settings as in \cite{millwood2023puf}. 
The VGG16 feature extractor was modified for classification tasks. We implemented it on a Raspberry Pi 3 Model-B, and it was observed that storing a model for three devices requires approximately 56 MB of memory, similar to the results in \cite{millwood2023puf}. Additionally, the time taken to perform the classification is around 5.72 seconds. We also simulate other functions on a single core 798 MHz CPU with 256MB of RAM, using the JCE library \cite{OracleJCA} to evaluate the execution time of the cryptographic primitives used in the proposed protocol. We show the results in Table \ref{tab:cost_op}. Finally, it is essential to highlight that within our proposed scheme, any IoT device has the capability to function as both the prover and verifier, with their computational costs being similar, as illustrated in Table \ref{tab:cost_op}.

% \begin{table}[h]
% \caption{EXECUTION TIME OF VARIOUS CRYPTOGRAPHIC Functions}
% \label{tab:cost}
% \begin{threeparttable}
% \begin{tabular}{p{1.2cm} p{0.5cm}cp{0.5cm}ccc}
%     \toprule
%     \textbf{Funtion} & $\mathrm{H}(\cdot)$ & $\mathrm{AEAD.Enc}(\cdot)$ & $\mathrm{DPUF}$ & $\mathrm{DPAN}$ & $\mathrm{KDF}$ & \textbf{Total}\\
%     \midrule
%     \textbf{Operation Times} & 2 & 2 & 2 & 1 & 1 & -\\
%     \midrule
%     \textbf{Time cost (ms)} & 0.026 & 2.9 & 0.12 & 5.72e3 & - & 5.8e3\\
%     \bottomrule
%     \end{tabular}
% \end{threeparttable}

% \end{table}

\begin{table}
\caption{Computation Cost for PhenoAuth}
\label{tab:cost_pro}
\centering
\begin{threeparttable}
\begin{tabular}{p{1.2cm} c}
    \toprule
    \textbf{Prover} & $2N_{\text{DPUF}} + 2N_{\text{H}} + 2N_{\text{AEAD.Enc}} + N_{\text{DPAN}} + N_{\text{KDF}}$\\
    \midrule
    \textbf{Verifier} & $2N_{\text{DPUF}} + 2N_{\text{H}} + 2N_{\text{AEAD.Enc}} + N_{\text{DPAN}} + N_{\text{KDF}}$\\
    \bottomrule
    \end{tabular}
\end{threeparttable}

\end{table}

\begin{table}
\caption{Execution time of the Various Operations Used in the Proposed Scheme}
\label{tab:cost_op}
\centering
\begin{threeparttable}
\begin{tabular}{ccc}
    \toprule
    \textbf{Operations} & \textbf{Prover} & \textbf{Verifier}\\
    \midrule
    $\mathrm{H}(\cdot)$(SHA-256) & 0.026ms & 0.026ms\\
    $\mathrm{AEAD.Enc}(\cdot)$ & 0.37ms & 0.37ms\\
    $\mathrm{DPUF}$ & 0.12ms & 0.12ms\\
    $\mathrm{DPAN}$ & 5.72s & 5.72s\\
    $\mathrm{HKDF}(\cdot)$(HMAC-based) & 2.9ms & 2.9ms\\
    \textbf{Total} & $\approx 5.7$ s & $\approx 5.7$ s\\
    \bottomrule
    \end{tabular}
\end{threeparttable}

\end{table}

% Then we use the JCE library to evaluate the execution time of the cryptographic primitives

\section*{Acknowledgement}

The work of Prosanta Gope was supported by The Royal Society Research Grant under grant RGS\textbackslash{}R1\textbackslash{}221183.  
Biplab Sikdar is supported by 
the National Research Foundation, Singapore and Infocomm Media Development Authority under its Future Communications Research Development Programme, under grant FCP-NUS-RG-2022- 019. Any opinions, findings and conclusions or recommendations expressed in this material are those of the author(s) and do not reflect the views of the National Research Foundation, Singapore and Infocomm Media Development Authority.

\section{Conclusion}

In this paper, we have introduced a novel PUF-based authentication protocol called  PhenoAuth for resource-constrained IoT devices using the concept of PUF-Phenotype. PhenoAuth has been designed to facilitate authenticated anonymous communication and ensure the integrity and privacy of IoT devices. One of the notable features of the proposed scheme is to detect any unauthorized attempted modification of the security credentials stored in the device memory.  
Moreover, the proposed PhenoAuth can also ensure backwards and forward security as long as the DRAM PUF remains uncompromised. One caveat, however, warrants discussion: PhenoAuth's reliance on the DRAM PUF, which is categorized as a weak PUF due to its finite responses (linearly growing with PUF size).  Future work into the effectiveness of reducing Phenotype size while maintaining high authentication performance would increase the number of available unique CRPs for use in this protocol. The protocol's sustainability, namely the number of supported sessions, is intrinsically linked to the key space, specifically, the stable portion of the DRAM PUF's response. In summary, PhenoAuth contributes a robust level of security for IoT device authentication. The adoption of PUF-Phenotype instead of traditional error correction algorithms affords the system greater adaptability in managing secret storage and addressing the intrinsic noise present in PUF responses.
% Though few responses are generated in \cite{millwood2023puf}, this is mainly due to large response sizes being used to generate the PUF Phenotypes.
% \todo[inline]{Limitations about key space}
\bibliographystyle{IEEEtran}
\bibliography{ref}

\end{document}

\title{Supplementary Material}
\maketitle

\begin{appendices}
\section{CRYPTOGRAPHIC NOTIONS}
\label{app:pre}
In this section, we define security primitives used in the protocol.

% \begin{definition}[DPUF]
% \label{app:dpuf}
%         Dram Physical Unclonable Function (DPUF) is a type of PUF that exploits the unique and unpredictable physical variations that occur naturally during the manufacturing process of DRAM to provide a device-specific and unclonable identifier.
% \end{definition}

% \begin{definition}[PUF Phenotype]
%     A PUF Phenotype is described as the full externally observable PUF behaviour, including noise, without supplementary knowledge of PUF structure, physical properties, or environment. Here we introduce two properties of PUF Phenotype:\\
%     \textbf{Classification Correctness:} The responses generated by corresponding DRAM PUF can be classified correctly, with the correct label and 
%     \textcolor{red}{1. Correct responses can be classified correctly with high probability. 2. Fake responses can not.}

%     \textbf{Unlearnable:} \textcolor{red}{The adversary can not learn anything from the trained model.}
    
% \end{definition}

\begin{definition}[AEAD]
    Authenticated Encryption with Associated Data. For such a cipher $\varepsilon = (E,D)$, the encryption algorithm works as follows:
    \begin{equation}
        c\gets E(k,m,d)
    \end{equation}
    where $c\in \mathcal{C}$ is the ciphertext, $k\in \mathcal{K}$ is the key, $m\in M$ is the message, and $d\in \mathcal{D}$ is the associated data. The decryption syntax is:
    \begin{equation}
        m/\bot \gets D(k,c,d)
    \end{equation}
    In this work, AEAD is presented as follows for for the sake of readability.
        \begin{equation}
    \begin{aligned}
        c &= \text{AEAD.Enc}( \underbracket{sk}_{session\ key}, \overbracket{N}^{counter}, \underbracket{\text{ad}}_{associated\ data}, m) \\
    \end{aligned}
    \end{equation}
    An AEAD cipher $\varepsilon = (E,D)$ is secure if it provides both CPA-secure and has ciphertext integrity.\\
    \textbf{CPA-secure:} An encryption scheme is CPA-secure if any polynomial-time adversary has a negligible advantage in distinguishing between the encryptions of any two chosen plaintexts.
    
    \textbf{Ciphertext Integrity:} A cryptographic scheme offers ciphertext integrity if it can be ensured that any alterations to the ciphertext (or associated data) post-encryption will be detected with high probability during decryption.

\end{definition}

\begin{definition}[Key Derivation Function]
A Key derivation function is an essential primitive in cryptographic systems that transfer a source of randomness into cryptographically strong secret keys, and sometimes the initial materials do not have to be distributed uniformly. According to the extract-then-expand approach, KDF has two components: extractors which extract a fixed-length key $K$ from entropy sources, random functions, which expand $K$ into several additional pseudorandom cryptographic keys. KDF takes four inputs: a random seed $r$, a length $L$, salt $s$ and context $c$, and returns a $L$-bits key $k$. The salt value is essential to obtain generic extractors and KDFs that can extract randomness from arbitrary sources with sufficiently high entropy. We capture the security of KDF by performing a game $\mathbf{Exp}_{KDF}$ between a challenger $\mathcal{C}$ and PPT $\mathcal{A}$:
\begin{enumerate}
    \item $\mathcal{C}$ generates a random value with distribution of $\alpha$. Then $\mathcal{C}$ generates a uniformly random salt $s$, and returns $(\alpha,s)$ to $\mathcal{A}$.
    \item $\mathcal{A}$ makes queries $(c_{i},l_{i})$ to $\mathcal{C}$, $\mathcal{C}$ returns $k_{i}=KDF(r,l_{i},s,c_{i})$.
    \item $\mathcal{A}$ generates $l$ and $c\notin (c_{1}\cdots c_{t})$ and sends them to $\mathcal{C}$, then $\mathcal{C}$ generates one bit $b\leftarrow \{0,1\}$ and computes $k_{0}=KDF(r,l,s,c)$ and $k_{1}\leftarrow \{0,1\}^{l}$. $\mathcal{C}$ returns $k_{b}$ according to $b$.
    \item $\mathcal{A}$ makes queries $(c_{i},l_{i}$ to $\mathcal{C}$ (where $c \notin c_{i}$), $\mathcal{C}$ returns $k_{i}=KDF(r,l_{i},s,c_{i})$.
    \item $\mathcal{A}$ outputs $b^{'}$.
\end{enumerate}
$\mathcal{A}$ wins the game if $b=b^{'}$, denoted as $\mathbf{Exp}_{KDF}=1$. We say the $KDF$ is secure if the advantage $\mathbf{Adv}_{KDF}(\mathcal{A})$ of $\mathcal{A}$ winning the game is negligible:
\begin{equation}
\mathbf{A d v}_{\mathrm{KDF}}(\mathcal{A})=2 \cdot\left|\mathbf{P r}\left[\left(b=b^{\prime}\right)\right]-\frac{1}{2}\right|
\end{equation}
\end{definition}

\begin{definition}[Collision-resistant Hash Function]
    A hash function $H$ maps input data of arbitrary length to fixed-size hash values, wherein it is computationally infeasible to find two different inputs $x$ and $x^{'}$ that result in the same hash value, i.e., $H(x) = H(x^{'})$. 

    \textbf{Preimage Resistance:} Given a hash value $h$, it should be computationally infeasible to find any input $x$ such that $H(x) = h$. In other words, it should be difficult to reverse-engineer the input from the given hash value.

    We capture the irreversibility of CRHF in the reverse attempt of $\mathcal{A}$. The advantage of breaking preimage resistance is the probability of $\mathcal{A}$'s successful reverse, $Adv_{CRHF}^{irv} = Pr(inverse(H(d))=d)$, where $d$ is randomly generated.

    \textbf{Second Preimage Resistance:} Given an input $x$, it should be computationally infeasible to find another input $x^{'} \neq x$ such that$ H(x) = H(x^{'})$. This property ensures that finding a different input with the same hash value is highly improbable.

    We capture the collision-resistance of CRHF in the colliding game of $\mathcal{A}$. The advantage of breaking second preimage resistance is the collision probability for randomly generated input, $\mathbf{Adv}_{CRHF}^{cls} = Pr(H(x)=H(x^{'})$.

    If the hash function $H$ satisfies \textbf{Preimage Resistance} and \textbf{Second Preimage Resistance}, we call $H$ a secure collision-resistant hash function.
\end{definition}

\begin{figure*}[t]
\centering
\begin{adjustbox}{max width=0.8\textwidth, max height=0.8\textheight}
\input{security_proof/Game_MU}
\end{adjustbox}
\caption{Security Reduction}
\label{}
\end{figure*}

\section{Security Model and Analysis}
\label{app:sec_ana}
\subsection{Adversarial Model}
\label{app:adversary model}
Based on the discussion in Section \ref{AM_informal}, we formally define the invasive adversary to analyse the security of proposed protocol. We allow the adversary to eavesdrop, modify, delay communication between the prover and the verified to gain the advantage of breaking the security. The adversary can also perform invasive operations to the on-chip NVM, which means he can readout the stored information in NVM. He is also allowed to modify the data, however, we guarantee that it makes no differences with blocking the entire communication channel (which can be detected easily), the authentication will fail and the best result for the adversary is the same as the deny of service (DoS) attack. For we assuming that the enrollment phase is performed in a secure channel, we consider $\adv$ can issue the following oracle queries:
\begin{enumerate}[label=$\cdot$]
    \item \textbf{Launch}$(1^{\lambda})$: A new session is started by the prover.
    \item \textbf{Send}$\prover(Dev_{id},m)$: Send arbitrary message $m$ to the device $Dev_{id}$.
    \item $\text{\textbf{Issue}}(DPUF)$: allows $\adv$ have physical access to the PUF instance $DPUF$, perform queries \textbf{Challenge}$(\cdot)$ and \textbf{Reveal}$(\cdot)$.
    \item $\text{\textbf{Reveal}}(C_{i}, \Delta_i)$: $\adv$ reveals sensitive data $C_{i}$ and $\Delta$ in NVM.
    \item $\text{\textbf{Corrupt}}(\cdot)$: allows $\adv$ modify the content in NVM.
    \item \textbf{Block}$(\cdot)$ allows $\adv$ block authentication massage $M_1,M_2$ for any side he wants in running of the protocol.
\end{enumerate}
We allow $\adv$ to have access to the NVM only when the session is not running on the device, and forbid it from the DPUF. Thus we give the definition for a \textbf{Cleanness predicate}.
\begin{definition}[Cleanness predicate]
    If the $\adv$ does not invoke \textbf{Issue}$(puf)$ during the whole game, and only invoke \textbf{Reveal}$(\cdot)$ and \textbf{Corrupt}$(\cdot)$ when the protocol is not running on the device, we call this session a \textbf{cleanness predicate}.
\end{definition}

\begin{figure*}[t]
\centering
\begin{adjustbox}{max width=0.8\textwidth, max height=0.8\textheight}
\input{security_proof/Game_privacy}
\end{adjustbox}
\caption{Privacy Game}
\label{appgame:privacy}
\end{figure*}

Apart from the cleanness predicate defined above, we also state which kind of interactions between adversary $\adv$ and challenger $\mathcal{C}$ are \textbf{clean}, in order to capture the nontrivial advantage $\adv$ can obtain without breaking the security of our protocol. Here we give the definition of matching.

\begin{definition} [Matching Session]
    In the D2D communication protocol between a prover $Dev_{p}$ and a verifier $Dev_{v}$, we say a session $\pi_{pv_i}$ is a matching session if device ID $Dev_{p}$ and $Dev_{v}$ are valid and the messages are honestly transferred between these two valid devices.
\end{definition}

\subsection{Security Model}
\label{sub:security_model}
Here we describe the security analysis model for the protocol. In the model, the challenger $\mathcal{C}$ maintains a group of IoT devices $D_{1},\dots,D_{n}$, with each device operating a number of instances of the proposed mutual authentication protocol $\Pi$. We note the the protocol performed between device $p$ and device $v$ as $\Pi_{pv}$, and the $i$-th session as $\pi_{pv_{i}}$. We define a matching session where the messages communicated between devices are honestly transferred. We capture the protocol's security $\Pi_{pv}$ through a game performed between a challenger $\mathcal{C}$ and an adversary $\adv$, noted as $\mathbf{Exp}_{\Pi,\adv}^{MA}(\lambda)$. The goal of $\adv$ is to model the communicated message $M_{1}:\{\alpha_{p_{i}},Tag_{p_{i}},AD_{p_i}\}$ and $M_{2}:\{\alpha_{v_{i}},Tag_{v_{i}},AD_{v_i}\}$ or derive the session key $mk$ used in $\Pi_{pv}$. At the end of the authentication process, $\mathbf{Exp}_{\Pi,\adv}^{MA}(\lambda)$ will output 1 (Accept) or 0 (reject). We say $\adv$ wins the game if it causes a \textbf{clean} session and $\mathbf{Exp}_{\Pi,\adv}^{MA}(\lambda)$ output 1, where $\mathcal{C}$ accepts the authentication without a matching session.  The basic game is considered as follows:

$\mathbf{Exp}_{\mathbf{\Pi,A}}^{MU}(\lambda)$:
\begin{enumerate}
    \item $(Dev_{id},C_{i}, \Delta_{i},pp)\xleftarrow[]{Random}\text{Setup}(\cdot)$;
    \item $(\hat{Dev_{id}},\widehat{M}_{1},\widehat{M}_{2},\widehat{mk_{i}})\xleftarrow[]{Random} \adv^{\text{Issue, Challenge, Reveal, Corrupt,Send$\prover$,Block}}((C_{i}, \Delta_{i},pp))$;
    % \item $S^',hd^'\xleftarrow[]{Random} \text{Reconfig}(\cdot)$
    \item $\Phi := Outcome(\widehat{mk_{i}},\widehat{M}_{1},\widehat{M}_{2})$;
\end{enumerate}

The challenger $\mathcal{C}$ executes a setup algorithm for enrolling into a trusted environment and all the public parameters $pp$ for initialization. Here $pp$ denotes all the available public parameters used to initialize the authentication, e.g. the length of security parameters and the setting up of $KDF$. The basic game of mutual authentication comprises two parts: prover authentication and verifier authentication. For both parts, we consider the same adversary model as discussed in Section \ref{app:adversary model}. We denote the advantage of the adversary $\adv$ in this game as $\advantage{MU}{\Pi,\adv}[(\lambda)]$, which is defined as the probability $\adv$ causes a winning event.

\subsection{Privacy Model}
\label{app:privacy}
Now we consider the indistinguishability-based privacy. In this case, the challenger $\mathcal{C}$ maintains 2 devices communicate with each other using the proposed protocol. The adversary $\adv$ eavesdrops somewhere in the network but tries to trace the communication parties. The privacy goal is that, in a clean session, the adversary $\adv$ cannot distinguish the true message between a randomly generated value. The basic game is considered as follows:

$\mathbf{Exp}_{\mathbf{\Pi,A}}^{\mathrm{IND}}(\lambda)$:
\begin{enumerate}
    \item $\adv$ issues \textbf{Launch}$(1^\lambda)$.
    \item $\mathcal{C}$ setup the system, start running the protocol $\Pi_{pv}$ on device $p$ and $v$, and obtain $M_1,M_2$ from the communication of the protocol.
    \item $\mathcal{C}$ combine $M1,M2$ as $m_0$, and then generate random value with the same length $|M_1|+|M2|$, noted as $m_1$. Then $\mathcal{C}$ generates a random choice $b$ from $\bin$, and then send $m_b$ to $\adv$.
    \item $\adv$ send his choice $b^*$ to $\mathcal{C}$.
\end{enumerate}
At the end of the game, $\mathcal{\mathbf{Exp}_{\mathbf{\Pi,\adv}}^{\mathrm{IND}}(\lambda)}$ outputs the result whether $b=b^*$, 1 as the same, 0 as not. We say $\adv$ wins the game if $\mathbf{Exp}_{\mathbf{\Pi,\adv}}^{\mathrm{IND}}(\lambda)$ outputs 1, and note the advantage that $\adv$ wins the game as $\advantage{\mathrm{IND}}{\Pi,\adv}=|\prob{(\mathbf{Exp}_{\mathbf{\Pi,\adv}}^{\mathrm{IND}}(\lambda)=1)}-\frac{1}{2}|$.

\subsection{Mutual Authentications}
We divide the analysis into two cases: prover impersonation and verifier impersonation. In prover impersonation, the protocol $\Pi$ accepts $M_1$ without an honest matching partner. We capture the security between a challenger $\mathcal{C}$ and adversary $\adv$. Figure \ref{game:MU} shows the whole game $\mathbf{Exp}_{\mathbf{puf,A}}^{\mathrm{MU}}(\lambda)$. Here, we show the analysis step by step, taking prover impersonation as an example.

\textbf{Prover Impersonation:}
Now, we consider the above model for analyzing the security of the proposed authentication protocol.

\textit{Proof:}  As discussed in Section \ref{sub:security_model} and shown in Figure \ref{game:MU}, the adversary $\adv$ tries to mimic an honest prover, by generating  $mk$ and $M_1$, without a corresponding matching session and also in a cleanness predicate. We start from the original game.

\textbf{MUGame 0:} This is the original game played between $\mathcal{C}$ and $\adv$. We denote the advantage of this game as $\advantage{MUGame0}{\Pi,\adv}$.

\textbf{MUGame 1:} In this game, $\adv$ replaces $mk_{p}\sample \bin^{|mk_{p}|}$. First, $\adv$ physically assesses a device before the session starts, and reads $Dev_{p},C_{i},\Delta_{i}$ from the NVM. Then it issues \textbf{Launch}$(1^\lambda)$ and tries to generate $M_1$. It should be noted that $\adv$ cannot access the NVM when the device is runnig. During the communication, $\adv$ eavesdrops on all the communicated information and stores it. After he collects the polynomial amount of data, he blocks the device's communication and tries to mimic it. For he collects associated data $\{AD\}$, the tags $\{Tag\}$ and the ciphertext $\alpha$, he invokes $\bdv_{1}$, which is built to break the security of AEAD algorithm. For the proposed protocol, updating the session key after every session, $\adv$ can obtain no advantage from knowing the old key for a new session; even $\bdv_1$ breaks the CPA-security of AEAD. The only chance is that, $\bdv_1$ can produce ciphertext and tag without knowing a correct key, which is bounded by the integrity game. Here, we introduce an abort event when the $AD_{p_i}$ is inconsistent with the tag $Tag$. Thus, the advantage after this game is:
\begin{equation}
    \advantage{MUGame1}{\Pi,\adv}<\advantage{MUGame0}{\Pi,\adv}+\epsilon_{1}+\advantage{AEADauth}{\bdv,\bdv_1}
\end{equation}

\textbf{MUGame 2:} In this game, $\adv$ treats $\Delta_{i}$ as the XORed responses from two DPUFs. For he already collects enough CRPs from the associated data in \textbf{Game 1}, he issues the adversary $\bdv_{2}$ and forwards the CRPs to it. Then $\adv$ can get the output $\widehat{R},\widehat{SR}$ from $\bdv_{2}$. $\adv$ invokes $KDF$ to derive the session key $\widehat{mk}$ from $\widehat{SR}$. Finally, as $\adv$ has the `session key' and associated data $AD_p$, he can generate $\widehat{\alpha_{p}},\widehat{Tag_p}$ using $AEAD.Enc(\cdot)$. The advantage of this game is:
\begin{equation}
    \advantage{MUGame2}{\Pi,\adv}\leq\advantage{MUGame1}{\Pi,\adv}+\advantage{ModPUF}{\Pi,\adv}
\end{equation}

We can find that, if the session key is secure, then $\adv$ can hardly break the security of authentication in the verifier impersonation game, if the $AEAD$ algorithm is secure, and $DPUF$ is secure against the modelling attack.

\textbf{Verifier Impersonation}
Now, we analyse the security of the proposed authentication protocol from the view of a prover. The goal is that, the adversary $\adv$ cannot cause a \textbf{clean} session and wins the $\mathrm{MU}$ without a matching session.\\
\textit{Proof:}  Similiar as discussed in prover impersonation case, the adversary $\adv$ tried to mimic an honest verifier by generating the session key $mk$ and $M_2$. In the proposed protocol, the prover and verifier store the same information on the device, and both allow the adversary to have physical access to the NVM when the protocol is not running. Because the transferred information $M_1$ and $M_2$ are highly similar, except the authentication state ($Auth_{req},Auth_{ok}$) and associated information $AD$. Thus the analysis is similar, that the challenger $\mathcal{C}$ will abort the game when the tag cannot be verified, or the protocol aborts itself as discussed in the protocol.

\subsection{Privacy Analysis}
Our goal concerning privacy is to make sure the adversary who eavesdrops in the network can not trace the identity of the communicated parties. In Section \ref{app:privacy}, we proposed the privacy model and the game $\mathbf{Exp}_{\mathbf{\Pi,A}}^{\mathrm{IND}}(\lambda)$. Here we will analyse step by step how the adversary can achieve the maximum advantage. From the protocol we know, to distinguish between different messages, $\adv$ can choose to trace $Dev_{id},\alpha,Tag, AD$.

\textit{Proof:} First we assume that our protocol holds security. Otherwise, the adversary can derive the session key $mk$ from $M_1$, then he can verify the $\alpha,Tag,AD$ from $M_1,M_2$ by encrypting the tag and associated data using $AEAD$ algorithm. Now we analyse the advantages step by step as shown in Figure \ref{appgame:privacy}:
\textbf{Game 0:} This is the original game, $\adv$ tries to trace the value of $\alpha$. We denote the advantage of this game as $\advantage{IND}{\Pi,\adv}$.

\textbf{INDGame 1:} In this game, $\adv$ tries to distinguish between the obtained $\alpha$ from $m_b$ and $\alpha_{blocked}$ from blocked $M_{{1}_{blocked}}$. The blocked message comes from the adversary's blocking query on a target device. He needs to categorise different messages by randomly guessing. Assuming there are $N_{session}$ sessions running at the same time, then he has a probability of $\frac{1}{N_{session}}$ to guess the correct matching information. $\adv$ issues another adversary $\bdv_{1}$ who can distinguish a ciphertext encrypted with $AEAD$ and a randomly generated value, without knowing the key $mk$. We note the advantage $\bdv_1$ can achieve as $\advantage{IND}{\mathrm{AEAD},\bdv_1}$. $\bdv_1$ will output $b1$, indicates whether they are encrypted using the same $mk$, $b1=0$ as the same, $b1=1$ as different. If $b1=1$, $\bdv$ will output $b^*=1$. Thus, the advantage after this game is:
\begin{equation}
    \advantage{INDGame1}{\Pi,\adv} < \advantage{INDGame0}{\Pi,\adv}+\frac{1}{N_{session}}\times \advantage{IND}{\mathrm{AEAD},\bdv_1}
\end{equation}

\textbf{INDGame 2:} In this game, $\adv$ tries to distinguish between the obtained $AD$ from $m_b$ and $AD_{blocked}$ from blocked $M_{{1}_{blocked}}$. The same as \textbf{INDGame 1} that $\adv$ needs to categorise different messages by randomly guessing. After obtaining them, $\adv$ issues another adversary $\bdv_{2}$ who can distinguish associated data $AD$ and a randomly generated value without knowing the key $mk$ and assessing the DPUF. We note the advantage $\bdv_1$ can achieve as $\advantage{IND}{\mathrm{AD},\bdv_1}$. $\bdv_1$ will output $b1$, indicates whether the $Dev_{id}$ are from the same device, or $\Delta^{1},\Delta^{2}$ are truly generated from the PUF's responses,$b2=0$ as yes and $b2=1$ as not. If $b2=1$, $\bdv$ will output $b^*=1$. Thus, the advantage after this game is:

\begin{equation}
    \advantage{INDGame2}{\Pi,\adv}<\advantage{INDGame1}{\Pi,\adv}+\frac{1}{N_{session}}\times \advantage{IND}{\mathrm{DPUF},\bdv_2}
\end{equation}

After $\adv$ performs the above operations, he sends the choice $b^*$ to the challenger $\mathcal{C}$. From the calculations we can find, if the $AEAD$ algorithm and $DPUF$ to be secure as claimed in Section \ref{app:pre}. Thus we prove the proposed protocol can guarantee privacy for both communication parties.

\end{appendices}